\documentclass[review]{elsarticle}

\usepackage{lineno,hyperref}
\usepackage[utf8]{inputenc}
\usepackage{natbib}
\usepackage{graphicx} 
\usepackage{amssymb} 
\usepackage{url}     
\usepackage{bm}      
\usepackage{color}
\usepackage{lipsum}
\usepackage{hyperref}
\usepackage{subfig}
\usepackage{float}

\newcommand{\na}{$^{22}$Na}
\newcommand{\gam}{$\gamma-$photons}

\journal{Nuclear Inst. and Methods in Physics Research, B}

\begin{document}
\begin{frontmatter}

\title{The SPOT-IL Positron Beam Construction and Its Use for Doppler Broadening Measurement of Titanium Thin Films}

\author[a]{P.~Or}
\author[a]{G.~Erlichman}
\author[a]{D.~Cohen}
\author[b]{I.~Sabo-Napadesky}
\author[b]{E.~Gordon}
\author[b]{S.~Cohen}
\author[c]{O.~Presler}
\author[d]{E.~O.~Cohen}
\author[d]{E.~Piasetzky}
\author[a]{H.~Steinberg}
\author[c]{S.~May-Tal~Beck}
\author[a]{Guy Ron \corref{corresponding}}
\cortext[corresponding]{Corresponding author email: Gron@phys.huji.ac.il}

\address[a]{Racah Institute of Physics, The Hebrew University, Jerusalem 91904, Israel}
\address[b]{Soreq NRC, Yavne, Israel 81800}
\address[c]{Physics department, NRCN, P.O.Box 9001, Beer-Sheva 84190, Israel}
\address[d]{School of Physics and Astronomy, Tel-Aviv University, Tel-Aviv, Israel 69978}

\begin{abstract}
The construction and first operation of the slow positron beam built at the Hebrew University is reported here. The beam follows a traditional design, using a $^{22}$Na source, a Tungsten moderator, and a target cell equipped with a load-lock system for easy sample insertion. The beam energy varies between $0.03$~keV and $30$~keV. The detection system consists of two high purity Germanium detectors, facing each other, allowing low-background Doppler-Broadening (DB) measurements. Event readout is done using a state-of-the-art compact desktop system. The target cell is designed to allow a combined measurement of DB and sample conductivity, with the flexibility to add more detection options in the future. The beam has been successfully tested by using it to charecterize Titanium (Ti) films. Two $1.2$~$\mu$m Ti films - as produced, and after annealing, were measured at various energies ($2$~keV - $25$~keV), and the results show consistent behavior with previous measurements.
\end{abstract}


\end{frontmatter}


\section{Introduction}

\par Positron Annihilation Spectroscopy (PAS) is a well established method used in the fields of solid state physics, chemistry, materials science and materials engineering \cite{dupasquier1995positron}. PAS harnesses the positron sensitivity to electron densities and defects in solid matrices for material characterization \cite{Tuomisto}. PAS techniques measure the photons emitted from the positron-electron  annihilation processes $(e^+ + e^- \rightarrow 2\gamma)$. In the center of mass system, the annihilation photons are emitted back to back, with energies equal to the rest mass of the positron (and electron) - $511$~keV, due to energy and momentum conservation. Since positrons lose their initial kinetic energy in matter before annihilating, the initial electron momentum affects the photons relative angle and their exact energy in the lab frame. The related spectroscopic measurements are Angular Correlation of Annihilation Radiation (ACAR) and Doppler Broadening (DB), respectively \cite{krause1999positron}. The latter is commonly used with positron beams as it is easier to implement. Another PAS technique, the Positron Lifetime (PL) measurement, is the most sensitive for defect identification, but requires time difference measurement between one of the annihilation photons and a "start" signal that can be obtained from a \na\ decaying source or a pulsed initial beam \cite{NISSILA2005778, doi:10.1002/pssc.200675745, BUTTERLING20112623}.
\par The origin of the positrons probing the material can be either a $\beta^{+}$ decaying isotope, or a positron beam. The former emits positrons with a wide energy distribution and with endpoint energies up to several MeV, having a large penetration range. It provides integrative information on bulk materials and efforts must be made to separate this material contribution from those of neighboring materials. Positron beams, on the other hand, provide mono-energetic beams, in a range of $\sim 0~-~50$~keV \cite{Tuomisto}. This allows control over positron penetration depths and characterization of materials `layer by layer' \cite{DB}, as well as measurements of thin films and foils. Additionally, the measurements are almost completely background free.
\par Positron beams have been developed since the late 80’s, the first ones being based on radioactive sources, such as \na\ \cite{Chilton_1995}. Today, they are in extensive use all over the world \cite{coleman2000positron, ZECCA20024, he2013}. High intensity beams are based on facilities, such as nuclear reactors \cite{falub2002magnetic, hugenschmidt2004neutron, hawari2009intense}, or linear accelerators~\cite{hughey1997experimental, hunt2002development, ebel1988slow,doi:10.1063/1.59156, Jungmann_2013}.
\par This paper presents the design and construction of a \na\ based positron beam at the Hebrew University of Jerusalem, together with results from the first measurements performed on Titanium (Ti) thin films, demonstrating its performance. The project was carried by a collaboration of researches from The Hebrew University, Tel-Aviv University, Nuclear Research Center (NRC) Soreq, and NRC Negev. This beam is part of the \textbf{S}low \textbf{PO}sitron facili\textbf{T}y in\textbf{ I}srae\textbf{L}, SPOT-IL, which also includes a compact life-time table-top system, discussed elsewhere \cite{SPOT-IL}. This facility is designed for basic and applied research. 
The SPOT-IL beam is equipped with two High Purity Germanium (HPGe) detectors, allowing low background Doppler Broadening (DB) measurements, in which the two \gam\ from the annihilation event are measured simultaneously. A special target cell was designed to allow combined electric conduction and annihilation measurements. We present beam design, the experimental setup of the DB measuring system, describe the Ti measurements, and discuss the results.


\renewcommand\floatpagefraction{.1}
\begin{figure}[hb!]
	\begin{center}
		\includegraphics[width=\textwidth]{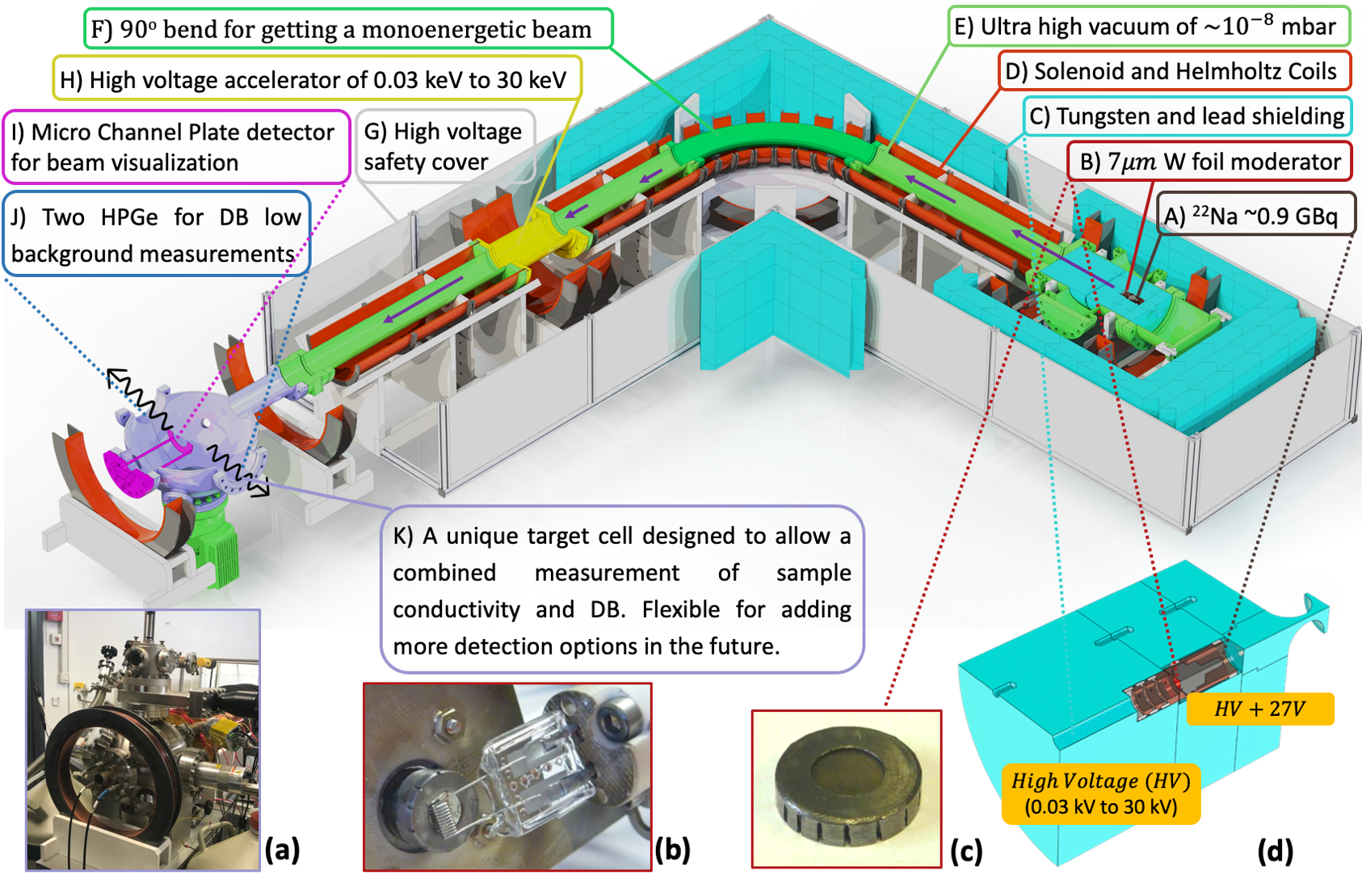}
        \vspace{-0.8cm}
        \caption{A model of SPOT-IL positron beam. [a] The target cell, equipped with a load lock system (from above), allowing insertion of samples to the center of the chamber without breaking the UHV. A web-camera looking through a small window at the MCP, allows beam visualization. The two HPGe detectors are mounted at close proximity to the sample, by two vacuum socket flanges.[b] The moderator and halogen lamp used for the annealing [c] a closeup to the W moderator attached to the source, with a diameter of $10$~mm [d] A section view of the source housing and pre-accelerator held at $27$~V over its surroundings which are held at high voltage.}\
		\label{model_fig}
	\end{center}
	\vspace{-0.8 cm}
\end{figure}

 
\section{Beam Construction}
 
\par A schematic of the slow positron beam is shown in Figure~\ref{model_fig}. The beam follows a traditional design~\cite{SPONSOR}: A \na\ positron source with current activity of $\sim0.9$ Gbq is used, in tandem with a $27$ V pre-accelerator unit, to inject the positrons into the beam-line. The positron system includes the source chamber, vacuum tubes surrounded by Copper coils producing a constant magnetic field, the accelerator unit, and a target chamber. The source chamber is placed on a High Voltage (HV) platform, which determines the energy of the positrons impinging on the sample in the grounded target chamber. The accelerator unit is comprised of a linear voltage divider using nine ceramic standoff rings. The complete beam-line is $\sim2$ m long, with a $90^{o}$ bend used to filter out high energy positrons. The entire system is held at a ultra high vacuum of $\sim10^{-8}$~mbar, providing for a long Mean Free Path ($MFP\gg2$ m) and preventing annihilation events before the positrons reach the target chamber and the sample.
\par The \na\ positron source was purchased from iThemba Labs \cite{Ithemba}. It is encapsulated in a dedicated housing \cite{na22source} with a $5~\mu$m thick Ti window, that only marginally affects the energy distribution of the emerging positrons \cite{SPONSOR}. A Tungsten (W) moderator was fabricated from a $7~\mu$m W foil welded to a Tantalum holder (see Figure~\ref{model_fig}). In order to achieve high efficiency the moderator was annealed prior to its attachment to the source housing. The annealing process included exposing the foil at close proximity to a W filament taken from a $36$~V $/ 400$~W halogen lamp, under $\sim28$~V (Figure~\ref{model_fig}). The moderator was first annealed for 5 minutes at $\sim2000^{o}$~C inside a vacuum chamber at $\sim10^{-7}$~mbar, and than for about 3 minutes at $\sim1300^{o}$~C under an oxygen atmosphere at $\sim10^{-4}$~mbar. \par After traversing the moderator foil a small fraction of the positrons is moderated to thermal energies and is emitted from the foil with E $\sim 3$~eV due to the W negative positron work function \cite{HUGENSCHMIDT2002283}. These positrons are directed into the beam-line using a pre-accelerator with an extra voltage of $+27$~V, applied by three commercially available $9$~V batteries. Then, they are selected by the $90^{\circ}$ bend and accelerated into the target chamber as a mono-energetic positron beam \cite{coleman2000positron}. Positrons emitted from the W moderator with higher energies do not bend as much and annihilate on the beam tube walls. Radiation shielding is provided by surrounding the \na\ source with a $55$~mm thick Tungsten cylinder inside the vacuum chamber (Figure~\ref{model_fig}). Outside the vacuum additional lead shielding walls, $50-150$~mm thick, surround the source chamber region and the bend region.
\par A detailed description of the beam-line magnetic fields design can be found in \cite{Beck_2014}. A constant magnetic field of $\sim9$~mT is applied along the beam-line, using solenoids surrounding the vacuum tubes and Helmholtz coils to keep that field in the vicinity of the source, accelerator unit and target positions. This field directs the positrons towards the target chamber and keeps them centered around the beam axis. A perpendicular pair of Helmholtz coils is located at the beam-line bend position, and another pair surrounds the target chamber, to allow control of the beam spot size and position at the target chamber. After the $90^{\circ}$ bend the slow positrons are accelerated using an accelerator unit, built from nine steel rings separated by ceramic rings which linearly ramps down the voltage to ground. The positron kinetic energy after the accelerator stage is equal to the voltage applied to the HV platform.
\par The beam-line is designed such that the target (sample) is kept at ground potential. This reduces the hazard of electrical shock at this readily accessible position. To allow the target to be kept at ground potential, all units upstream of the acceleration unit have to be kept on high-voltage platform. This means that the moderator assembly and the pre-accelerator unit, as well as the bending section, should all be kept at high-voltage. Because of the uniform base voltage, the pre-acceleration of low energy positrons and their selection by the bend are indifferent to their absolute value. The HV platfrom may be set between $0.03$~kV and $30$~kV and is shielded using a Perspex cage. 
\par The grounded target cell in Figure~\ref{model_fig} is designed to allow a combined measurement of sample conductivity and DB, with the flexibility to add more detection options in the future, such as low temperature for integrated in-situ electronic measurements. It also includes a load-lock system for insertion and removal of samples to and from the target chamber without breaking vacuum.

\section{Experimental Setup}

\par At the center of the target chamber, just behind the sample position, a Micro Channel Plate (MCP) detector with  a phosphor screen is installed. The MCP allows visualization of the beam spot and is monitored by a simple web-camera through a small window. By applying a small magnetic field perpendicular to the positron flux, the beam spot position on the screen is easily controlled.
\par Two High Purity Germanium detectors (HPGe) are installed outside the target chamber, facing each other at an angle of $180^{\circ}$, for low background Doppler Broadening (DB) measurements. Data acquisition and readout is done by a compact desktop system - CAEN dual digital multi channel analyzer (DT5780SC) \cite{DT5780} - serving as a HV power supply and a preamplifier. The module is equipped with two independent $16$~k digital multi-channel analyzers, and is able to manage coincidences between the two HPGe detectors, allowing easy subtraction of background events.
\par The HPGe detectors energy spectra were calibrated using $^{137}Cs$ and $^{60}Co$ $\gamma-$\ sources with energy peaks at $661.7$~keV, $1173.2$~keV, and $1332.5$~keV. The energy resolution (FWHM) of the $661.7$~keV and $1173.2$~keV peaks were measured to be $1.3$~keV and $1.6$~keV respectively. Since energy conservation requires that the two $\gamma$s total energy is twice the positron (or electron) rest mass, $511$~keV , by taking into account only events where the energy sum of the two $\gamma$s is twice that value, namely $E=1022$~keV, background events are easily suppressed.
\par In a DB measurement two parameters are defined for the estimation of the broadening of the $511$~keV  peak - the Shape (S) and Wing (W) parameters. The first indicates the percentage of annihilation events close to the $511$~keV peak (within a range of $509.9$~keV $\leq E_{\gamma} \leq 512.1$~keV), the second indicates the percentage of annihilation events far from the peak (with a range of $507.3$~keV~$\leq E_{\gamma} \leq 509.3$~keV and $512.7$~keV $\leq E_{\gamma} \leq 514.7$~keV) \cite{SW}. In metals, the S and W parameters relate to conducting electrons and bound electrons, respectively, and thus carry information on defects populations (S) and their chemical neighbors (W). The ratio between them is constant for defects of the same kind, even if their concentration varies \cite{krause1999positron}.
\par
\begin{figure}[t]
	    \subfloat[]{\includegraphics[width=5.7cm]{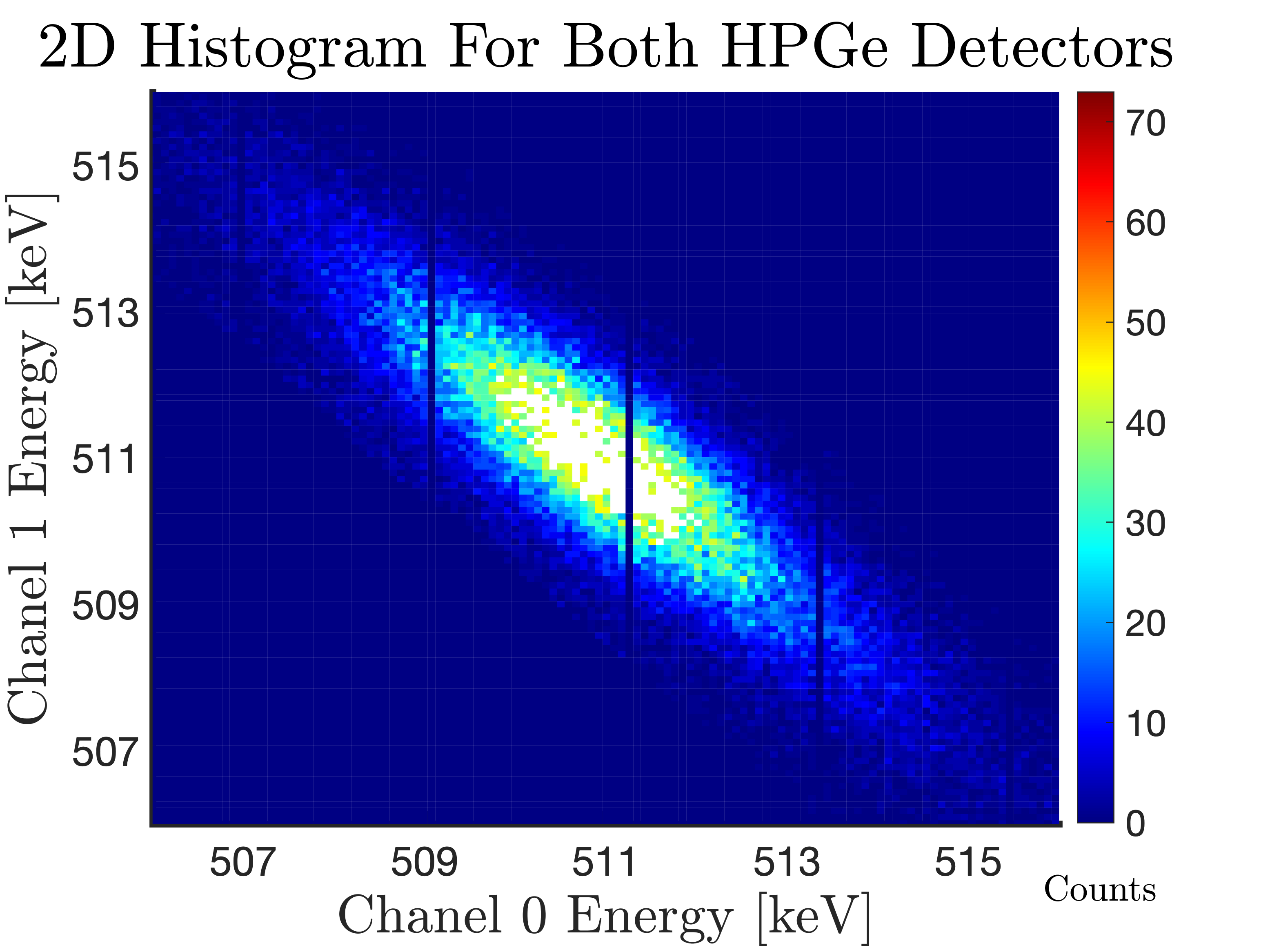}}%
        \qquad
        \subfloat[]{\includegraphics[width=5.7cm]{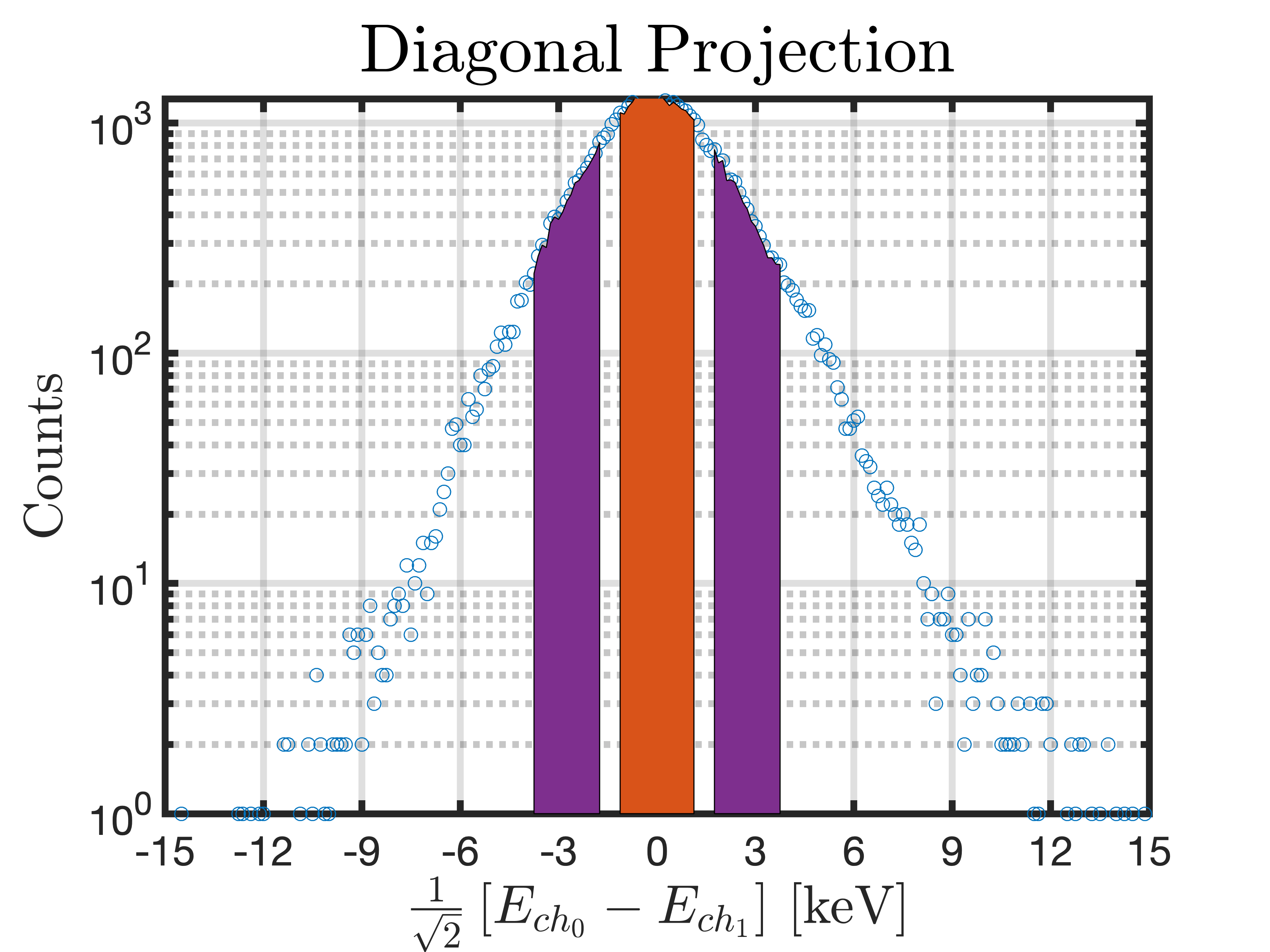}}%
        \caption{(a) A 2D energy histogram of all detected events which their energy sum is $1022$~keV~$\pm~2.6$~keV. (b) A typical diagonal projection of the 2D histogram, in log scale. The red and purple areas are used to extract the S and W parameters respectively.}\
        \label{2D_diag_fig}
    	\vspace{-0.8cm}
\end{figure}

 A typical 2D spectrum measured by both detectors is shown in Figure~\ref{2D_diag_fig}. Only events that occur simultaneously within a $1~\mu$s window are recorded. In the offline analysis, only events of $2\cdot511\pm2.6$~keV are taken into account, with $2.6$~keV being $3\sigma$ of the Gaussian around $E_{Ch0}+E_{Ch1}=1022$~keV, corresponding to $99.7\%$ of the events. The diagonal projection of the 2D histogram in Figure~\ref{2D_diag_fig} is used to extract the S and W parameters. A typical projection is shown in Figure~\ref{2D_diag_fig}.


\renewcommand\floatpagefraction{.1}%


\section{Ti measurements}


\renewcommand\floatpagefraction{.1}
\begin{figure}[t]
	\begin{center}
		\subfloat[]{\includegraphics[width=6cm]{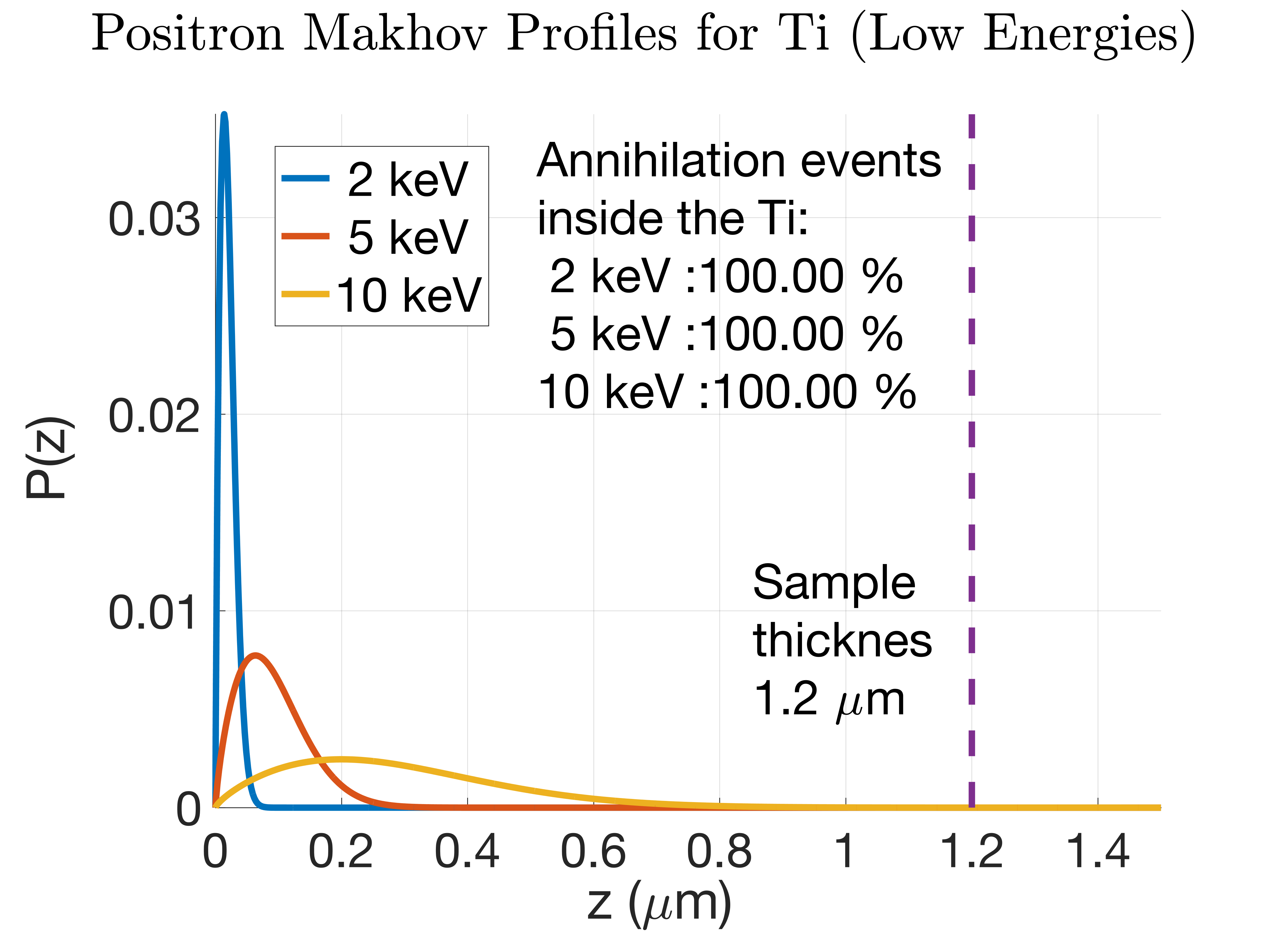}}%
        \subfloat[]{\includegraphics[width=6cm]{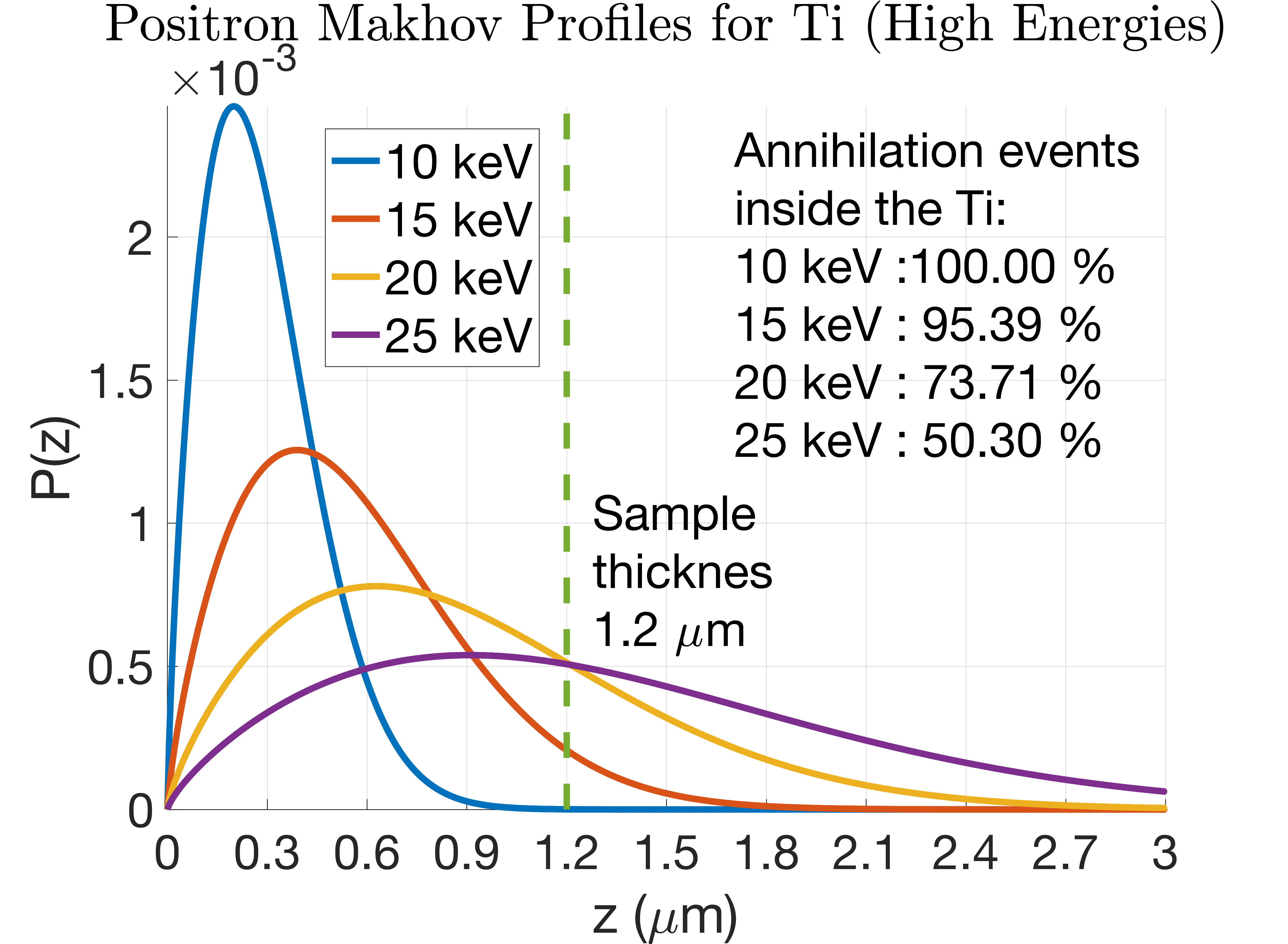}}%
		\caption{(a,b) Makhov profiles for $1.2~\mu$m thick Ti samples, and the fraction of positrons annihilating in it, as a function of their initial energy. It can be easily noticed that for energies below $10$~keV , the majority of positrons are annihilating inside the Ti sample, and as the energy increases, a larger fraction of positrons are annihilating in the Cu substrate.}\
		\label{Makhov}
		\vspace{-0.8cm}
	\end{center}
\end{figure}


\par The beam was commissioned using measurements on two Ti thin film samples. The samples were deposited on commercial copper OFHC ($50 \times 50 \times 0.25$~mm sized) substrates by a DC Pulse magnetron sputtering system. Before deposition, both substrates were immersed in Ammonium persulfate $2\%$ to remove oxides and then cleaned with deionized water, Acetone, Petroleum ether, and dried with Nitrogen. The substrates were inserted to a sputtering system, with pure ($99.95\%$) Ti cathode, in two different batches. In both processes, after reaching chamber base pressure of $10^{-7}$~mbar, the substrates were outgassed to elevated temperature ($\sim450^{\circ}$~C) and then cooled to $250^{\circ}$~C for two hours of deposition. During processes, the cathode power was $50$~W / $-100$~V bias and Argon partial pressure $6~\mu$bar (system pressure $5\cdot10^{-3}$~mbar). One of the samples was annealed to $450^{\circ}$~C for one hour immediately after deposition, inside the sputtering system.
\par Both samples were measured at six different positron energies ($2$~keV, $5$~keV, $10$~keV, $15$~keV, $20$~keV, and $25$~keV). An estimation of the positron implantation depth was done using Makhov profiles \cite{DRYZEK20084000}, as seen in Figure~\ref{Makhov}. For positron energies $<10$~keV, all positrons annihilate within the $1.2~\mu$m Ti films. For higher energies there is an increasing probability for the positrons to also annihilate in the Cu substrate.
\par
\begin{figure}[H]
	\centering
    {\subfloat[]{\includegraphics[width=6cm]{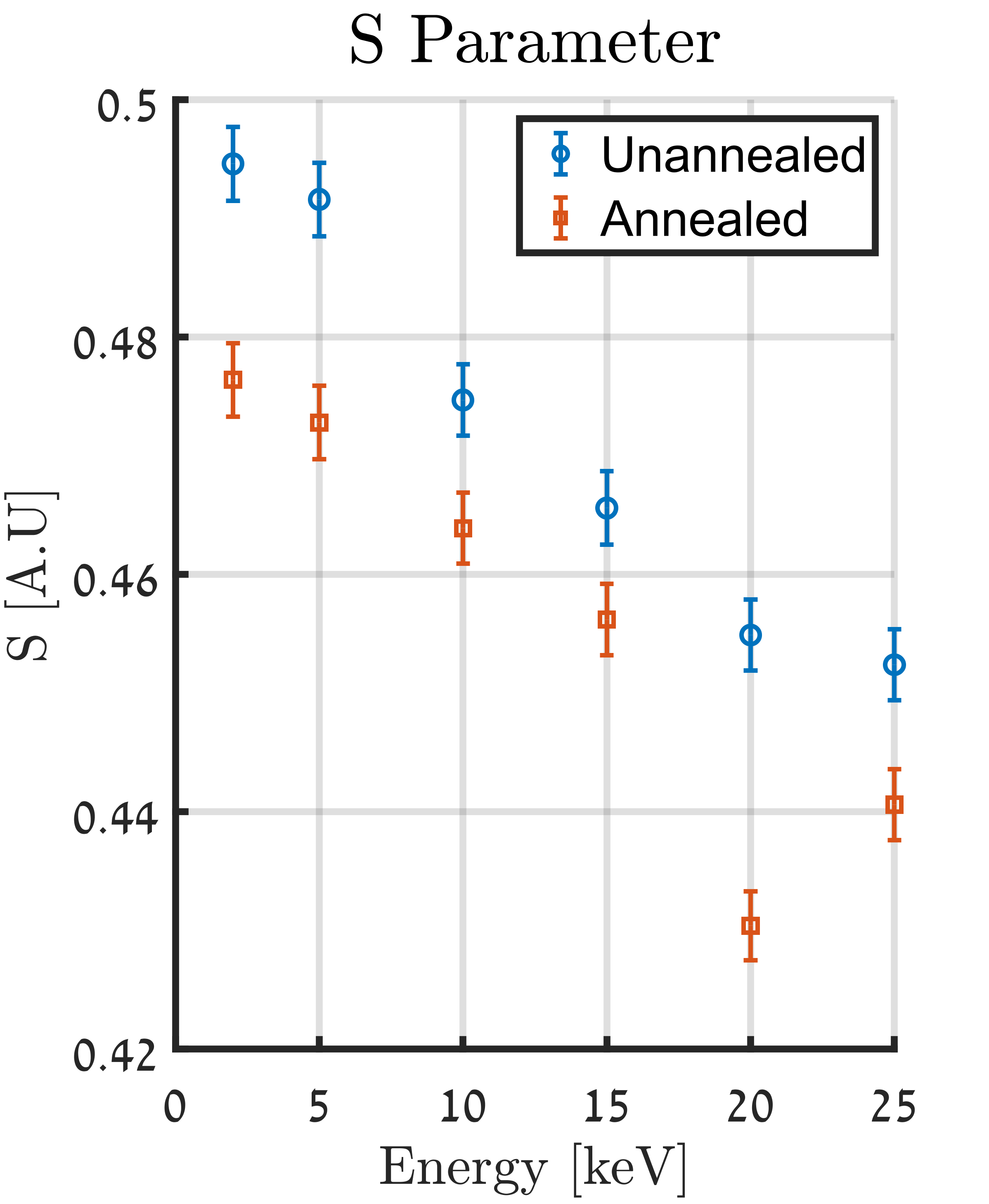}}}%
    {\subfloat[]{\includegraphics[width=6cm]{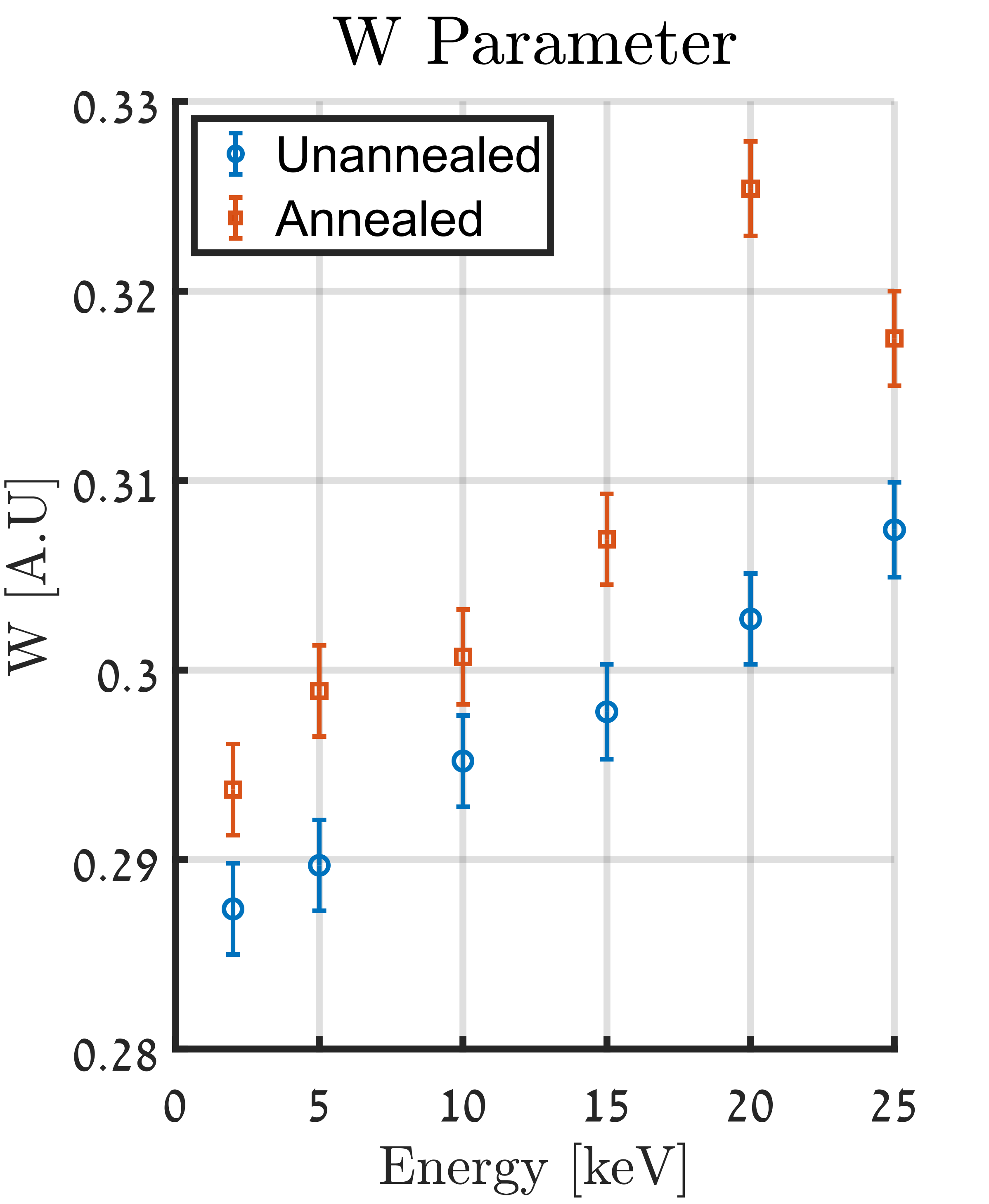}}}%
    
    {\subfloat[]{\includegraphics[width=12cm]{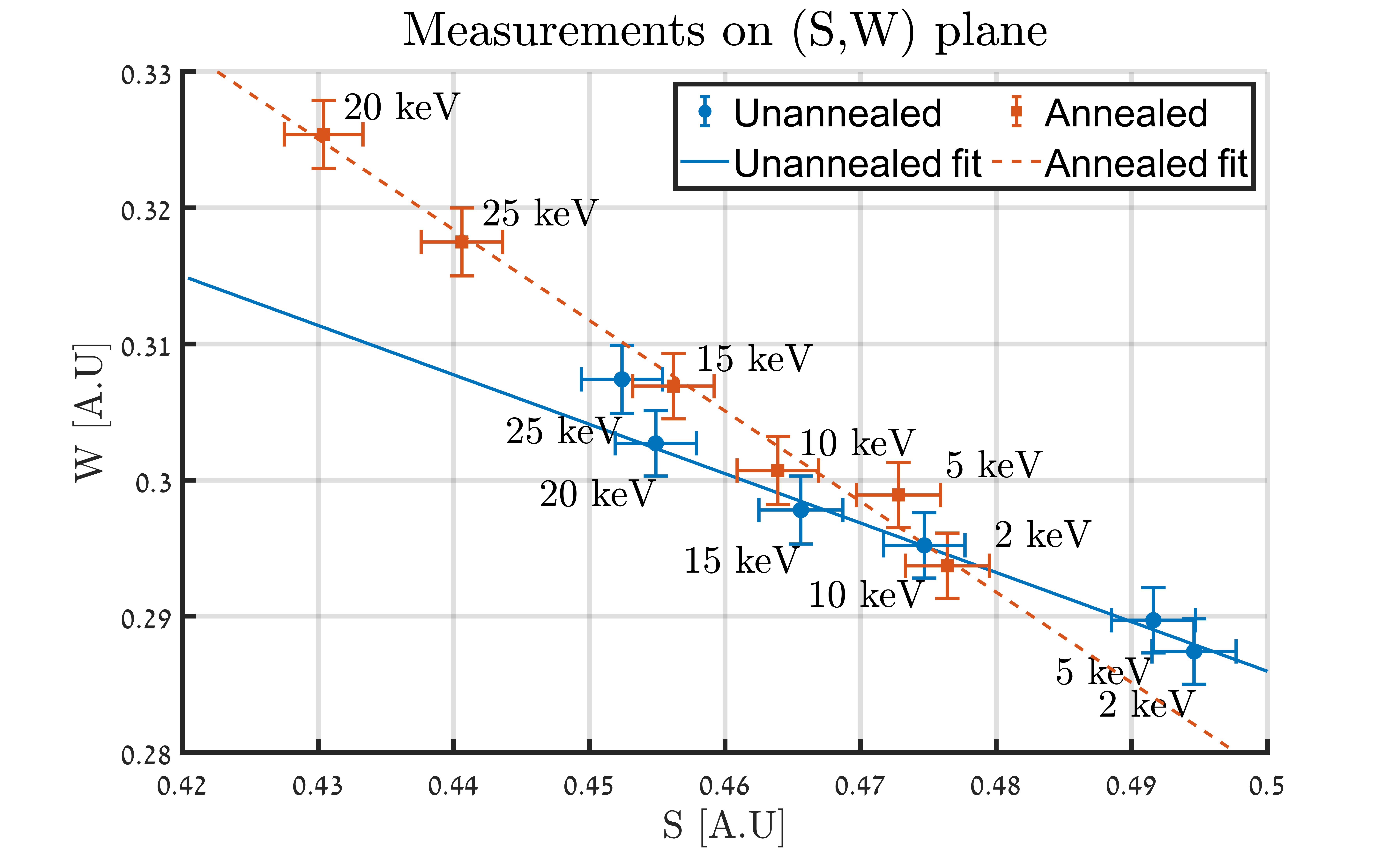}}}
    \caption{Two Ti thin films, annealed and unannealed, are compared for different initial positron energies. Measurements of S (a) and W (b) as a function of the energy, respectively. (c) Measurements presented on (S,W) plane, each measurement is color-coded and labeled with the measurement energy. The unannealed sample shows higher S values evidence of more defects inside the Ti bulk, as expected.}\
    \label{SW_fig}
\end{figure}

 A measurement in each energy was taken with $\sim50$~k significant events (passing the requirement of  $2\cdot511\pm2.6$~keV), over $1\sim2$ days. For each energy, S and W parameters were extracted and plotted on (S,W), (S,E), and (W,E) planes, as shown in Figure~\ref{SW_fig}. The S and W parameter values for the two samples change significantly through the Ti thin film - from surface to bulk. With $15$~keV positrons, there is already a small contribution from annihilations in the Cu substrate, which increases with increasing positron energy, affecting the total S and W values. Since the Cu substrate has been annealed with the Ti film, the (S,W) values for these substrates should not necessarily be the same. A comparison of the two samples reveals higher S values for the unannealed sample as compared to the annealed one and lower W values, indicating more defects in the as-deposited Ti film on the Cu substrate. As energy increases and positrons are implanted in the bulk, S values decrease and W increases, implying higher defect concentration close to the surface of the films.
\par For the high energy points, where the Cu layer already contributes to the results, few features are clearly seen: (1) the (S,E) and (W,E) points at $20$~keV for the annealed sample deviates from the general trend of all other measured points for this sample, (2) the $25$~keV point tends to approach the values of the unannealed sample, (3) both $25$~keV points show different behavior on the (S,W) plane as compared to other points of the same samples. Fit lines to the data points up to $20$~keV are shown in Figure~\ref{SW_fig}. The different slopes for the unannealed and annealed Ti films, show the annealing effect. In fact, the temp. used for degassing probably already annealed the Cu substrate and thus, the $25$~keV (S,W) point of the unannealed sample approaches the annealed line. The transition from $20$~keV to $25$~keV at the annealed sample presents a reverse trend in the (S,W) parameters behaviour, may imply the positrons annihilate in a mixed Ti/Cu diffusion layer that is formed close to this boundary, containing less defects.


\renewcommand\floatpagefraction{.1}


\par Within measurement uncertainty, the (S,W) plot Figure~\ref{SW_fig} shows a constant slope for the unannealed sample, indicating the same type of defects exist throughout it, with varying concentration. The annealed sample shows a significantly different slope, indicating different defects. The smaller S parameter in the annealed sample indicates also smaller defects, and one can conclude that the annealing process has decreased the defect size. This picture is in good agreement with previous Ti thin film measurements \cite{MISHEVA199626}.
\par These results show feasibility of the beam operation, performance of the detection system, and correct analysis. More work is needed in order to increase detection rate for faster data collection, such that more data points will be measured, improve the analysis details and to measure separately Copper substrates treated the same way as with the Ti films on top of them for more precise results.

\section{Summary}

\par The slow positron beam constructed in the Hebrew University of Jerusalem was successfully commissioned, using measurements on two Ti $1.2$~$\mu$m thin films, showing behaviour consistent with previous measurements. Control of the positron energies and implantation depth was demonstrated, and the detector system was shown to be capable of preforming low background DB measurements, using a compact desktop digital system.
\par Our beam is designed to combine DB measurements with sample conductivity. Previous research in our group showed the development of Graphene-based positron charge sensor \cite{doi:10.1063/1.5053477}, and we intend to continue this research using the slow positron beam. Currently, the beam configuration allows DB measurements at room temperature only, but simple modifications of the system will allow measurements under in-situ cooling or heating.


\section*{Acknowledgments}

This project was supported by the Ministry of Economy under the KAMIN program and by the Pazy grant of the IAEC. P.Or was supported by the Ministry of Science and Technology ,Israel. We wish to thank the SPONSOR group at HZDR, Germany, for sharing their knowledge and for guiding and supporting us in all aspects of the source and beam design. Special thanks to Wolfgang Anwand from HZDR, who was always available and willingly devoted his time to teach and help us, and to R. Krause-Rehberg, F. Deininger from University Halle, and A. Wagner from HZDR who helped us with the moderator annealing process.

\bibliography{References.bib}

\end{document}